\documentclass[%
reprint,
superscriptaddress,
longbibliography,
 amsmath,amssymb,
 aps,
pra,
]{revtex4-2}
\usepackage{amsmath,amssymb,amsfonts,mathrsfs}
\usepackage{graphicx}
\usepackage{dcolumn}
\usepackage{bm}
\usepackage{multirow}
\usepackage{hhline}
\usepackage{longtable}
\usepackage{xy}
\usepackage{bm}
\usepackage{cancel}
\usepackage[usenames,dvipsnames,svgnames,table]{xcolor}
\usepackage{epstopdf}
\usepackage{float}
\usepackage{ulem}
\usepackage{soul}

\def\cdd{C_{\mathrm{dd}\,}}
\def\Ldip{L_{\mathrm{dip}}}

\begin{document}


\title{Mutual Dipolar Drag in a Bilayer Fermi Gas}

\author{B. Renklioglu}

\email{brenklioglu@sisecam.com}%
\address{Department of Physics, Bilkent University, Ankara 06800, T{\"u}rkiye}
\address{ Science, Technology and Design Center, Şişecam, Istanbul, T\"{u}rkiye}
\author{Ben Yu-Kuang Hu}
\email{byhu@uakron.edu}
\affiliation{Department of Physics, University of Akron, Akron, Ohio 44325-4001, USA}
\author{M. \"{O}. Oktel}
\email{oktel@fen.bilkent.edu.tr}
\affiliation{Department of Physics, Bilkent University, Ankara 06800, T{\"u}rkiye}
\author{B. Tanatar} %
\email{tanatar@fen.bilkent.edu.tr}
\affiliation{Department of Physics, Bilkent University, Ankara 06800, T{\"u}rkiye}

\begin{abstract}
We consider two-dimensional spin-polarized dipolar Fermi gases confined in a double-layer system and calculate 
the momentum transfer between the layers as a function of temperature to investigate the transport properties of 
the system. We use the Hubbard approximation to describe the correlation effects and the screening between 
the dipoles within a single layer. The effective inter-layer interaction between the dipoles across the layers is obtained 
by the random-phase approximation (RPA). We calculate the interaction strength and the layer separation distance
dependence of the drag rate, and we show that there is a critical distance below which the system is unstable. 
In addition, we calculate the typical behavior of the collective modes related to the density fluctuations.

\begin{description}
\item[PACS numbers]
\end{description}
\end{abstract}

\maketitle


\section{Introduction}
In the last two decades, transport properties of two-dimensional (2D) electron and hole systems have attracted a great deal of interest as a result of the unique nature of the temperature dependence of resistivity which shows insulating (metallic)
behavior at low (high) densities \cite{kravchenko}. In this context, the inter-layer resistivity has been measured for many
systems such as
2D electron systems in $\mbox{AlGaAs/GaAs}$ double quantum wells~\cite{G1,G3,Solo,Eis} and electron-hole bilayer
structures~\cite{Sivan}. The characteristics of the inter-layer resistivity are determined by the Coulomb scattering
so the drag measurement is regarded as an efficient probe to study the properties of the intra- and inter-layer
electron-electron interactions in low-density bilayer systems~\cite{Jauho, Hu, Rojo,C_drag}.

On the other hand, in another area of physics, studies on ultracold atoms have provided a huge amount of information about the unique properties of ultracold systems which include atomic species with magnetic dipole moments (e.g. Chromium (Cr) atoms~\cite{cr1,cr2,cr3,cr4,cr5,cr6,cr7}, Erbium (Er) atoms~\cite{Er,Er2,Er3,ErDy} and Dysprosium (Dy) atoms~\cite{ErDy,Dy,Dy2,Dy3}) and polar molecules with electrical dipole moments~\cite{ref1,ref2,ref3,ref4,ref5,ref6,ref7,ref8,ref9,ref10}. 
Crucially, ultracold gases of fermionic atoms and molecules with strong dipolar interactions have been realized experimentally \cite{DipoleFermi1,DipoleFermi2,DipoleFermi3,DipoleFermi4}.
Understanding the distinct nature of these
polar atoms and molecules is very important because they exhibit novel phases and previously unexplored regimes~\cite{Goral,Cooper,Sansone,Pollet,Demler}.

In our previous studies~\cite{cool1,cool2}, three of us implemented a variant of the Coulomb drag phenomenon,
namely "mutual dipolar drag", to a model in which a contactless heat transfer occurs through dipolar coupling in the
linear (without an external field) and non-linear (with an external field) regimes. We investigated the applicability and
efficiency of this sympathetic cooling method for the cooling of ultracold dipolar gases. We concluded that even
for the most magnetic dipolar atomic gases, the typical optical lattice length scale is too large a separation between
the layers for significant drag-like effects to be observed. Thus, our results indicated that ultracold molecules with
three orders of magnitude stronger dipolar interaction were the only possibility for observing such effects.
While there has been significant progress in creating ultracold dipolar molecules, these systems are still much more
fragile compared to the ultracold atomic systems. So far, no experiment has even measured transport properties,
let alone interlayer drag in molecular ultracold gas.

However,  a remarkable recent experiment observed sympathetic cooling without contact in a Dy gas.~\cite{Ketterle}
This experiment overcame the limited strength of atomic dipolar interaction between the layers not by enhancing the
dipole strength but by making the layers closer together. The 50\,nm separation between the two layers was obtained
using a dual polarization and frequency scheme. A ten-fold decrease in the separation of the layers enhances the
inter-layer excitation by three orders of magnitude, leading to the observation of contactless drag-like effects.
This experiment creates new impetus to probe the interlayer physics of double-layer systems, complementing
the physics observed in bilayer electron systems.

Momentum transfer between dipolar gasses has previously been investigated by Matveeva, Recati and
Stringari\cite{nonDiss}. Their work establishes that typical experimental lengths and time scales of dipolar
gases are suitable for the detection of the dipolar coupling between the two layers, particularly for dipolar molecules.
As the focus of Ref.~\onlinecite{nonDiss} is on the coupling of the center of mass motion between the two layers,
interactions are treated within the Hartree approximation, and any dissipative effects resulting from particle-hole
excitations in each layer are ignored. However, as we have investigated in the case of thermal coupling between
the layers these particle-hole excitations are the dominant mechanism for equilibration between the layers and
give a local mechanism for momentum transfer independent of the size of the clouds. While the coupled oscillations
of the center of mass of the clouds are dependent on the geometry of the clouds, the damping of these oscillations is controlled by the rate of momentum transfer between the layers by forming particle-hole excitations.

In this paper, we focus on the transport properties of  2D dipolar Fermi gases in analogy to a similar effect
observed in electronic systems. We consider two parallel layers of 2D spin-polarized dipolar Fermi gases
separated by a distance $d$, in which there is no inter-layer tunneling. The mutual dipolar drag
(related to the momentum transfer) is calculated as a function of temperature between the layers of the system
in which the particles interact only with the long-range interactions. In our investigation of the interaction strength
and the layer-separation distance dependence of the drag rate, we find an instability in the system that is
analogous to one described in Ref.\,\onlinecite{Akaturk}.  We use the Hubbard approximation to define the correlation effects between dipoles within a single layer. For the effective interlayer interactions, we adopt the random-phase approximation (RPA). In addition, the transport characteristics of the 2D dipolar Fermi gases indicate a typical behavior of the collective modes related to the charge-density fluctuations.

In the literature, studies show that there is both energy and momentum exchange in bilayer Fermi systems
as a result of the Coulomb drag~\cite{C_drag}. In contrast to the bilayer electronic systems, dipolar interaction
is the dominant long-range interaction in ultracold systems. Our previous work~\cite{cool1,cool2} indicated
that heat transfer depending on the mutual dipolar drag, can be used as an efficient cooling method for ultracold gases.
According to our mechanism, the layer-separation distance should be kept constant around the magnitude
of the dipolar length scale of the system. Specifically, the required interlayer distance is consistent with the
typical feature size of the potential in current experiments for ultracold polar molecules, so this mechanism
can be set up without much difficulty. We have shown that the equilibration time constants between the layers
for different polar molecules are of the order of tens of milliseconds, which are smaller than typical trap lifetimes.
Based on our findings in this study, we believe that it would be of interest to develop the corresponding model
using dynamic optical lattices and investigate the transport properties of ultracold systems.

The rest of the present paper is organized as follows: In the next section, we introduce our model in detail
and outline our approach. In Section III, we present the results of our calculations in various parameter regimes.
Section IV contains the discussion and conclusion of our work. Appendix A gives details of the derivation of the
dipolar drag rate and Appendix B presents details of the collective mode dispersions.

\section{The Model and Method}
\begin{figure}
\includegraphics[scale=0.7]{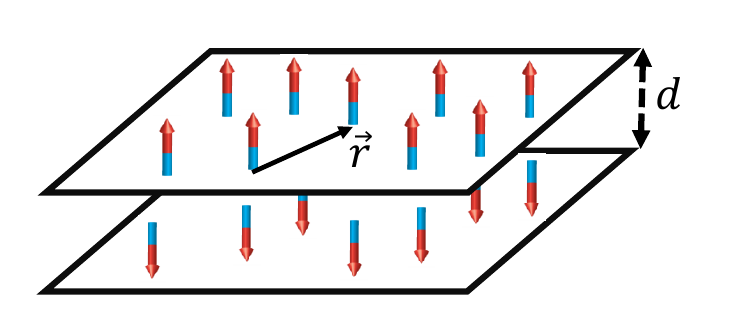}
\caption{Schematic view of the system. We consider two dipolar Fermi gas layers with equal densities at thermal equilibrium. Dipoles are oriented perpendicular to the layers but antiparallel in two layers. Separation distance is $d$.}\label{model}
\end{figure}

In this study, we consider a two-dimensional spin-polarized dipolar Fermi gas, confined in two parallel layers
separated by a distance $d$, as shown in Fig.~\ref{model}. The system is in thermal equilibrium and there is
no tunneling between the layers. The dipoles are polarized perpendicular to the planes, but the relative direction
of this polarization is antiparallel in two layers (see, Fig.~\ref{model}).~\cite{Akaturk}
The bare intra-layer (within a single layer) $V_{11}$ and inter-layer (across the layers) $V_{12}$ are given by
\begin{equation}\label{V11r}
V_{11}(r) = {V}_{22}(r) = \frac{C_{dd}}{4\pi}\frac{1}{r^3}\,,
\end{equation}
and
\begin{equation}\label{V12r}
V_{12}(r)=\frac{C_{dd}}{4\pi} \frac{r^2-2d^2}{(r^2+d^2)^{5/2}}\,,
\end{equation}
where the indices $1$ and $2$ denote different layers and $r$ indicates the in-plane distance between dipoles.
$C_{dd}$ is the dipole-dipole coupling constant, which is $C_{dd}=\mu_0\mu^2$ for magnetic dipole moments
$\mu$, and $C_{dd}=p^2/\epsilon_0$ for electric dipole moments $p$. Here, $\mu_0$ is the vacuum permeability,
$\epsilon_0$ is the permittivity of free space. The inter-layer interaction $V_{12}$ is repulsive for small $r$
but attractive for large $r$. Attractive interaction may lead to pairing but it is either absent or
extremely weak in the antiparallel configuration we choose.~\cite{fedorov, boudjemaa,attractint}

The Hamiltonian of the system is described by
\begin{equation}\label{Hamil}\begin{split}
\mathcal{H}=&-\frac{\hbar^2}{2m}\sum_i \left(\nabla_{1i}^2+\nabla_{2i}^2\right) + U_1(r_{1i})  + U_2(r_{2i}) \\
            &+\frac{1}{2}\sum_{i,j} \Big[V_{11}(|\vec{r}_{1i}-\vec{r}_{1j}|)+V_{22}(|\vec{r}_{2i}-\vec{r}_{2j}|)\Big]  \\
            &+\sum_{i,j} V_{12}(|\vec{r}_{2i}-\vec{r}_{1j}|)  \,,
\end{split}\end{equation}
where $m$ is the mass of the particles, $U_1$ and $U_2$ are the box potentials with a certain width that confine the particles in the direction perpendicular to the layers \cite{boxtrap} and the sums are carried out over the particles in each respective layer.  
We assume negligible widths for simplicity as the finite width effects will only soften the interaction potentials without making qualitative changes in the results reported here.

We define the following characteristic length scales. (i) $a_0=C_{dd}m/(4\pi \hbar^2)$, which is a measure of the 
strength of the dipole-dipole interaction. (ii) The inner dynamics of the system are determined through the Fermi energy
($E_F=\hbar^2 k_F^2/2m$) by the average distance, $k_F^{-1}$, between two fermions within a layer 
(or the density of a layer $n$). Here, $k_F=\sqrt{4\pi n}$ is the Fermi wave number for a
spin-polarized system, and $n$ is the 2D density of a single layer. 
(iii) The distance $d$ between the layers indicates the corresponding geometry of the model.

\subsection{Effective Interactions}

We use the Hubbard approximation to obtain an effective intra-layer interaction in Fourier space without
any cut-off parameter. We assume the dipoles have charge $\pm e$ which are at $z=\pm \Ldip/2$ where
$\Ldip$ is the physical size of the dipole.  Then, the bare intra-layer interaction is given by
\begin{align}
\displaystyle V_{ii}(q) &= \frac{e^2}{\epsilon_0 q}\left[1-\exp(-q L_{\mathrm{dip}})\right]\nonumber\\
&= \frac{\cdd}{\Ldip^2 q} \left[1-\exp(-q \Ldip)\right]\nonumber\\
&= \frac{\cdd}{\Ldip} - \frac{\cdd}2 q + O(\Ldip)\\
&=V_0- \frac{\cdd}2 q.
\end{align}
Here, $i$ indicates one of the layers and $V_0$ is the cut-off parameter related to the width of layers~\cite{Li,Sarma,Zinner}. Note that the $V_0$ diverges as we take $\Ldip\rightarrow 0$.

The only place where $V_{ii}$ occurs in this work is in the screening of the intra-layer interaction, as $V_{\mathrm{eff},ii}= V_{ii}(q) [1 - G(q)]$ where $G(q)$ is the intra-layer local field factor. We use the Hubbard approximation of the local field factor~\cite{V,vignale}, $G(q) \approx V_{ii}(\sqrt{q^2 + k_F^2})/V_{ii}(q)$, which gives
\begin{equation}
V^{(H)}_{ii}(q)=\frac{\cdd}2 \left[\sqrt{q^2 + k_F^2} - q\right].\label{V11}
\end{equation}
In the above, we took the Laurent expansion form of  $V_{ii}(q)$ in powers of $\Ldip$.  This effective interaction
$V_{ii}^{(H)}(q)$ contains the intra-layer correlation effects to a certain extent and it has been widely used
in electronic systems~\cite{vignale}. It also renders the inter-layer interactions free from the parameter $\Ldip$.
It is possible to go beyond the Hubbard approximation by including
higher order correlation effects \cite{Abed}.

The inter-layer interactions are treated within the random-phase approximation (RPA) hence the Fourier transform
is given by
\begin{equation}\label{V12}
V_{12}(q)=-\frac{C_{dd}}{2} q \exp(-qd)\,.
\end{equation}
As the separation $d$ between the layers should not be too close to avoid tunneling this should be a
reasonable approximation.

If the finite width effects of the layers need to be
considered, one has to integrate out the confinement
wavefunction in the $z$-direction before taking the 
two-dimensional Fourier transform of the intra- and 
inter-layer interactions given in Eqs.\,(1) and (2), respectively.
This procedure will bring out form factors $F_{ij}(q,L)$
to modify the interactions given in Eqs.\,(6) and (7)
where $L$ is the well width. Such modifications have been
taken into account in the case of harmonic confinement. \cite{Zinner,Kopietz}
It would be interesting to consider the dimensionality
effects on the drag rate by tuning the harmonic
confinement in different directions to produce cigar and
pancake shaped Fermi vapors.
In our case, finite width effects should only make qualitative changes.
The Coulomb drag rate in double quantum well systems
has been shown to be qualitatively similar for different layer widths. \cite{vazifehshenas}

\subsection{Mutual Dipolar Drag}
 In a double-layer system, when an external disturbance is applied to one of the layers, a particle current flows
 through the ``active'' layer with a current density. As a result of the inter-layer interaction across the layers,
 an induced disturbance appears in the ``passive'' layer. This is the mutual dipolar drag effect and it can be
 quantified by the drag rate $\tau_D^{-1}$. The drag rate can be described as the net average rate of momentum
 transferred to each particle in the passive layer, per unit drift momentum per particle in the active layer\cite{Hu}
\begin{equation}\label{defn3}\begin{split}
\tau_D^{-1} =\frac{\overline{(\partial p_2/\partial t)}}{\overline{p_1}} \,,
\end{split} \end{equation}
where the overbar indicates an ensemble average and $p_i$ is the momentum per particle in the corresponding layer. To investigate the transport properties of the system, we derive the drag rate between the layers, given by Rojo~\cite{Rojo} when the inter-layer interaction is treated perturbatively~\cite{Sivan}. In Appendix A, the corresponding derivation is presented, according to this calculation the momentum transfer is given by
\begin{equation}\label{result}\begin{split}
\tau_D^{-1} =\frac{\hbar^2 }{8m_1n_2k_BT\pi^2} &\int_0^\infty dq \ q^3\ |W_{12}|^2\ \\
 &\times\int_0^\infty d\omega \frac{\mbox{Im}\chi_1(q,\omega)\mbox{Im}\chi_2(q,\omega)}{\sinh^2(\hbar\omega/2k_BT)} \,.
\end{split} \end{equation}
Here, the dynamically screened effective interaction is indicated by $W_{12}=V_{12}/\varepsilon(q,\omega,T)$
where $\varepsilon(q,\omega,T)$  is the temperature-dependent total screening (dielectric) function obtained from the
random phase approximation (RPA)~\cite{Hu} as
\begin{equation}\label{epsilon}\begin{split}
\varepsilon(q,\omega,T)= \\
          & \left[ 1-V_{11}^{{(H)}}(q)\chi(q,\omega ,T)\right]\left[ 1-V_{22}^{{(H)}}(q)\chi(q,\omega ,T)\right] \\
                    & -\left[V_{12}(q)\right]^2\chi(q,\omega, T)\chi(q,\omega ,T)\, ,
\end{split}\end{equation}
where $\chi(q,\omega,T)$ is the temperature-dependent non-interacting density-density dynamical
response function of a single layer.
In this study, we focus on a symmetric system in which the densities of the layers are equal $n_1=n_2)$.

\section{Results}
In our presentations below, we shall use the following dimensionless quantities: $Q=q/k_F$,
$\Omega=\hbar\omega/E_F$,
$D=dk_F$, $t=k_BT/E_F$, $\tilde{\mu}=\mu/E_F$, $\tilde{\chi}=(\pi\hbar^2/m)\chi$, and $\lambda=a_0k_F$.

The intra-layer interaction for the Hubbard approximation is
\begin{equation}\label{V11d}
\mathbb{V}_{11}^{(H)}(Q)= \mathbb{V}_{22}^{(H)}(Q) = \left (\frac{m}{\pi\hbar^2}\right ) 2\lambda \left[\sqrt{Q^2+1}-Q\right]\,,
\end{equation}
and the inter-layer interaction is given by
\begin{equation}\label{V12d}
\mathbb{V}_{12}(Q)=- \left (\frac{m}{\pi\hbar^2}\right )2 \lambda Q \exp\left(-D Q\right)\,.
\end{equation}
In terms of these parameters, the dimensionless drag rate is obtained as
\begin{equation}
\tilde \tau_D^{-1} = \frac{1}{\pi t} \int_0^\infty dQ\ Q^3 \int_0^\infty d\Omega\vert\widetilde{W}_{12}\vert^2\frac{[\mbox{Im}\tilde\chi(Q,\Omega)]^2}{\sinh^2(\Omega/2t)}
\end{equation}
where $\tilde \tau_D^{-1} = \frac{\hbar\tau_D^{-1}}{E_F}$ and $\widetilde{W}_{12} = (\pi\hbar^2/m)\mathbb V_{12}/\varepsilon(Q,\Omega)$\, ,

\subsection{Analytic Calculations}

We analytically investigate the drag rate dependence on $\lambda$ and $D$ in the
limit of zero temperature ($T\rightarrow 0$) while the effective intra-layer interaction is obtained by using the
Hubbard approximation. For small values of the temperature, we use the small $\Omega$ expansion of $\chi$
(because the integral is cut off by the $\sinh^2$ term for $\Omega\gg t$), which for the spin-polarized case is
\begin{subequations}
\begin{align}
\mbox{Im}\chi(Q,\Omega) &\approx \frac{m^2 \omega}{4\pi \hbar^3 q k_F} = \frac{m}{8\pi\hbar^2} \frac{\Omega}{Q}\, ,\\
\mbox{Re}\chi(Q,\Omega) &\approx -\frac{m}{2\pi \hbar^2}.
\end{align}
\end{subequations}
Also, in this case we can use the static limit of $\varepsilon$,
\begin{align}
\varepsilon(Q,\Omega= 0) &\approx (1 + \lambda \tilde V^{(H)}_{11}(Q))^2 - \lambda^2 \tilde V_{12}^2(Q),
\end{align}
where the dimensionless intra-layer and inter-layer interactions are
$\tilde{V}^{(H)}_{11}(Q)= \left (\frac{\pi\hbar^2}{m\lambda}\right )\mathbb{V}_{11}^{(H)}(Q) = \mathbb{V}\sqrt{Q^2+1}-Q$
and $\tilde{V}_{12}(Q)= \left (\frac{\pi\hbar^2}{m\lambda}\right )\mathbb{V}_{12}(Q) = Q\exp{(-QD)}$, respectively.  Henceforth, for convenience we will drop the superscript $(H)$ in $\tilde V^{(H)}_{11}$.

Putting all these into the expression for the scaled drag gives
\begin{equation}\begin{split}\label{abc}
\tilde\Gamma_D & = \frac{\tau_D^{-1} \hbar E_F}{(k_BT^2)} \\
                      & = \frac1{\pi t^3}\int_0^\infty  d\Omega \ \frac{\Omega^2}{\sinh^2(\Omega/2t)}  \\
&\ \times\int_0^\infty dQ\ Q \frac{\lambda^2 \tilde V_{12}^2(Q)}{\Bigl[(1 + \lambda \tilde V_{11}(Q))^2 - \lambda^2 \tilde V_{12}^2(Q)\Bigr]^2}\,.
\end{split} \end{equation}
Note that the second ($Q$) integral is independent of the temperature $t$.  Using the relation
\begin{equation}
\int_0^\infty \frac{t^s}{\sinh^2(t)}\ dt = \frac{\Gamma(s+1)}{2^{s-1}\,}\zeta(s)\quad \mbox{for $\mbox{Re}[s] > 1$}
\end{equation}
in the first ($\Omega$) integral in Eq.~(\ref{abc}) gives
\begin{equation}\begin{split}
\int_0^\infty d\Omega \ \frac{\Omega^2}{\sinh^2(\Omega/2t)} & = 8t^3\int_0^\infty dx\ \frac{x^2}{\sinh^2(x)} \\
                                                            & = 8t^3 \zeta(2) = \frac{4\pi^2t^3}3.
\end{split} \end{equation}
Thus, Eq.~(\ref{abc}) becomes
\begin{equation}\begin{split}
\tilde\Gamma_D = &\frac{4\pi}3 \lambda^2 \int_0^\infty dQ\ Q^3\ \exp(-2QD)\   \\
                    &\times \frac{1}{\Bigl[(1+\lambda(\sqrt{Q^2 + 1} - Q)^2 - \lambda^2 Q^2\; \exp(-2QD)  \Bigr]}\nonumber\\
\end{split} \end{equation}
To simplify this expression, we introduce a new variable $y=DQ$ so that
\begin{equation} \begin{split}
\tilde\Gamma_D= &\frac{4\pi\lambda^2}{3D^4} \int_0^\infty dy\ y^3 \exp(-2y)\    \\
                    &\times \frac{1}{\left[\left(1 + \frac{\lambda}{D}\left\{\sqrt{y^2 + D^2} - y\right\}\right)^2 - \frac{\lambda^2}{D^2} y^2 \exp(-2y)  \right]^2}.
 \end{split}\end{equation}
Since this expression is temperature independent, it implies that $\tau^{-1} = \tilde\Gamma_D (k_B T)^2/(\hbar E_F)$ grows quadratically with temperature for small $t$.

For the weak coupling limit ($\lambda/D  \ll 1$), the denominator can be replaced by $1$ (no screening) in which case
\begin{equation}\label{weak}
\tilde\Gamma_D \approx \frac{4\pi\lambda^2}{3D^4} \int_0^\infty dy\ y^3\;\exp(-2y) = \frac{\pi\lambda^2}{2 D^4}.
\end{equation}
On the other hand, for the strong coupling limit ($\lambda/D  \gg 1$), we can rewrite the integral as
\begin{equation}
\tilde\Gamma_D
= \frac{4\pi}{3\lambda^2} \int_0^\infty \frac{dy\,y^3 \exp(-2y)}{\displaystyle \biggl[\left(\frac{D}{\lambda} + \sqrt{y^2 + D^2} - y\right)^2 - y^2 \exp(-2y)  \biggr]^2}.\label{strong}\\
\end{equation}
Note that in terms of physical length scales, $\lambda/D = a_0/d$.

For large $\lambda$, the integral in Eq.~(\ref{strong}) approaches a constant and therefore $\tilde\Gamma_D \propto \lambda^{-2}$.
For sufficiently large $\lambda$, as $D$ (the scaled inter-layer distance) is decreased, at a critical value $D_c(\lambda)$
the denominator of the integrand in Eq.~(\ref{strong}) vanishes at some value of $y$, causing the integral to diverge.
This corresponds to the development of a density wave instability in the coupled system, which is analogous to the
instability described in Ref.~\cite{Akaturk} for bilayers of antiparallel dipoles as shown in Fig.\,5 in that reference. 
At large $\lambda$, $D_c(\lambda\rightarrow\infty) \approx 1.077$.  For $D<D_c$, the system is no longer homogeneous, and therefore our model is not expected to describe the drag rate accurately.

The long-range and anisotropic character of the dipolar interaction results in some fundamental instabilities toward pattern formation. Recently, roton instability in bosonic dipolar gases has been found to cause supersolid states, which are stabilized by quantum fluctuations \cite{Roton1,Roton2,Roton3}.  The instability arising in our calculation has two fundamental differences from the roton instability. It requires two separated layers, similar to the two-component gas experiments, and it would be stabilized by the Pauli pressure due to the fermionic statistics of the system.  Thus, the mechanism is closer to a spin-wave instability rather than a roton instability.

When $D$ is fixed, the low temperature scaled drag rate $\tilde\Gamma$ initially increases when $\lambda$ is increased from zero because the effect of screening is small, and therefore increasing $\lambda$ increases the coupling between the layers, and hence the momentum transfer rate.  When $\lambda \approx D$, the screening of the inter-layer interaction starts to dominate, and further increasing $\lambda$ decreases the inter-layer coupling and hence the momentum transfer rate.

\subsection{Numerical Calculations}
The dimensionless drag rate between the layers is numerically calculated as a function of temperature $t=k_BT/E_F$ for weak ($\lambda/D \ll 1$) and strong ($\lambda/D \gg 1$) coupling limits of the system by using dynamic and static screening separately.

\begin{figure}[h]
\includegraphics[scale=0.55]{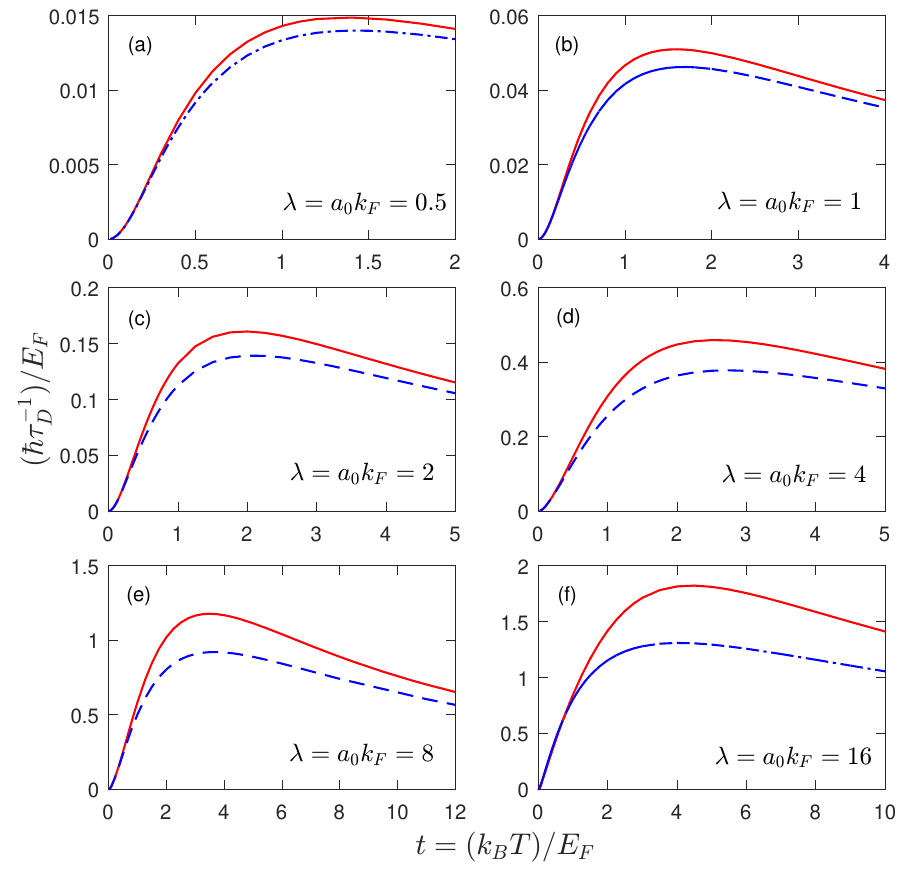}
\caption{(color online) The dimensionless drag rate $\tau_D^{-1}$ is plotted as a function of temperature $t=k_BT/E_F$
for different values of the interaction strength $\lambda=0.5, 1.0, 2.0,4.0, 8.0\; \mbox{and}\; 16.0$ when the layer
separation distance is kept constant at $D = dk_F=1$. In each graph, the momentum transfer is calculated by using
the dynamical (red solid line) and static screening (blue dashed line) separately. Here, the effective intra-layer
interaction is obtained by using the Hubbard approximation.}\label{small_hub}
\end{figure}

We obtain the drag rate as a function of temperature for different values of interaction strength
$\lambda=0.5, 1, 2, 4, 8\; \mbox{and}\; 16$ as shown in Fig~\ref{small_hub}.
We use the Hubbard approximation to obtain the intra-layer interactions in the system.
As we increase the interaction strength $\lambda$, the amount of transferred momentum increases as expected
(see Fig.~\ref{small_hub}). In this figure, the dimensionless distance is constant at $D=dk_F=1$.
While this value of $D$ is less than $D_c(\lambda\rightarrow\infty) \approx 1.08$, for all these values of
$\lambda$ used, $D > D_c(\lambda)$ [see Fig.~1A in Ref.~\onlinecite{Akaturk}],
indicating that the system is in the homogenous liquid regime.  In addition, as a result of plasmon enhancement,
the difference between the drag rates obtained from the dynamical (shown by the red solid line in Fig.~\ref{small_hub})
and static (indicated by the blue dashed line in Fig.~\ref{small_hub}) screening also increases as the value of the
dipolar coupling increases.

\begin{figure}[h]
\centering
\includegraphics[scale=0.55]{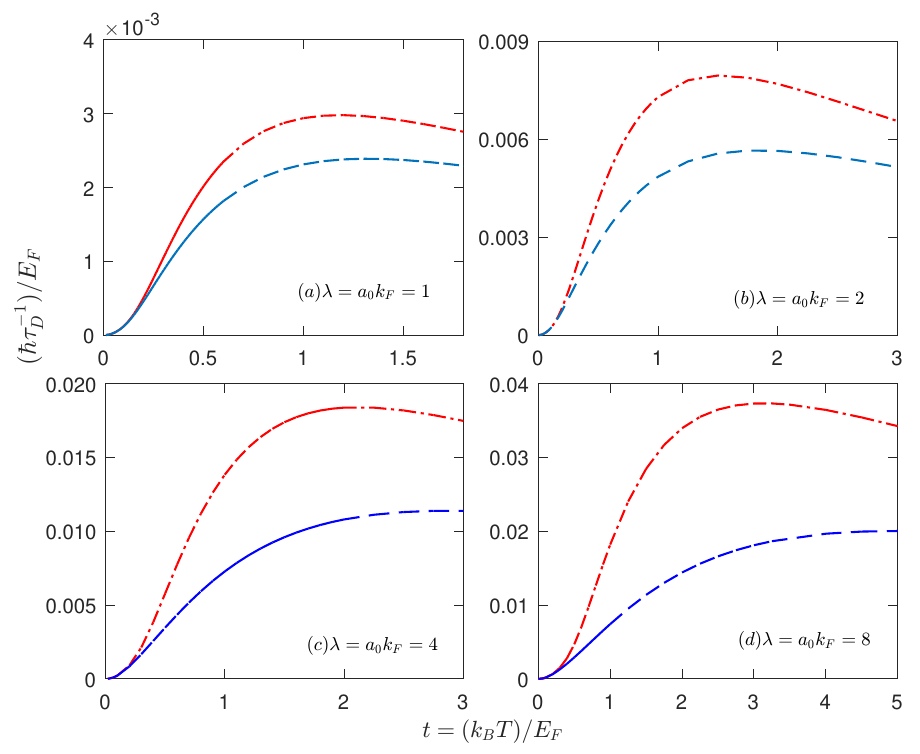}
\caption{(color online) The dimensionless drag rate is obtained as a function of temperature for various coupling
strengths ($\lambda=1,2,4,8$). In all of the plots, the static and dynamic screening for the system are considered
separately (indicated by blue-dashed lines and green-solid lines, respectively). Here, the dimensionless layer
separation distance is constant at $D=dk_F=2$. }\label{d2_Hubb}
\end{figure}

To understand the layer separation distance dependence, we also obtained the dimensionless drag rate as a function of temperature when the dimensionless distance between the layers is kept constant at $D=dk_F=2$, as shown in Fig.~\ref{d2_Hubb}. The momentum transfer rates are calculated for both dynamic and static screening cases. In all of these plots, the system is in a homogenous liquid state ($D=2>D_c$).  Comparing Figs. \ref{small_hub} and \ref{d2_Hubb} one sees, as expected, that the drag rate decreases as the distance between the layers increases.  Figs. \ref{small_hub} and \ref{d2_Hubb} also show that as $\lambda$ increases, the peak in the drag rate as a function of temperature is pushed to a higher temperature, and the magnitude of drag at this peak increases with $\lambda$.

\begin{figure}[h]
\centering
\includegraphics[scale=0.55]{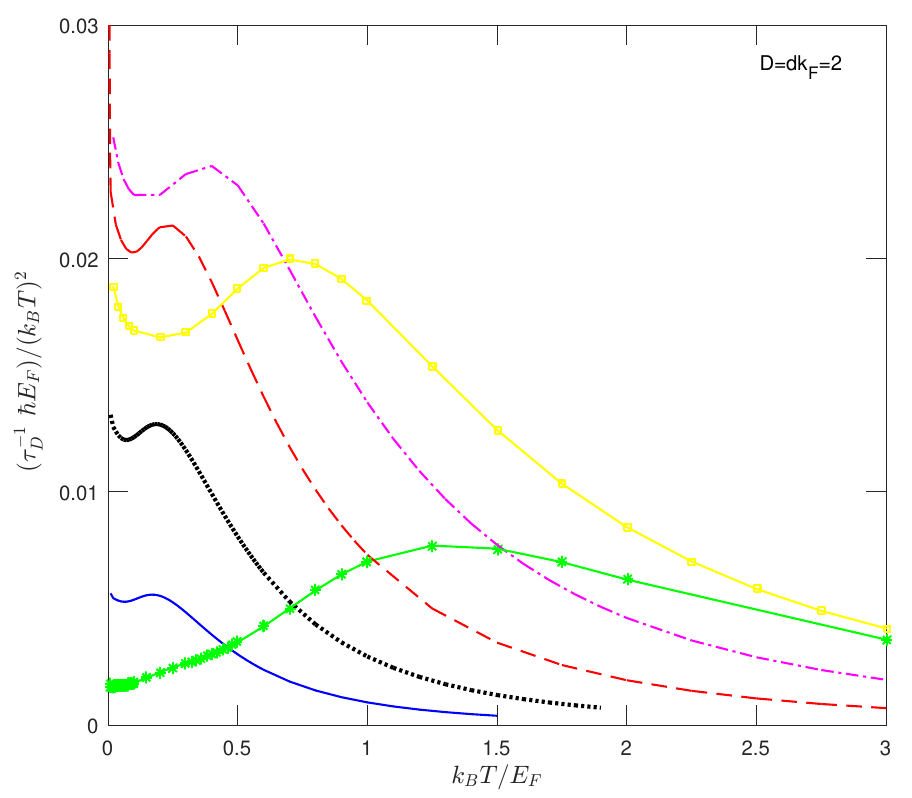}
\caption{(color online) Temperature dependence of the drag rate scaled by $T^{-2}$ for different values of the interaction strengths while the layer separation distance is constant at $D=dk_F=2$. The solid (blue), dotted (black), dashed (red), dashed-dotted (pink), solid with square (yellow), and starred-solid (green) lines are obtained using the dynamical screening for $\lambda=0.5,1,2,4,8\mbox{ and }16$, respectively. Here, the Hubbard approximation is adopted for the effective intra-layer interactions, $V_{11}^{(H)}$.  Note that the $T\rightarrow 0$ scaled drag rate increases as $\lambda$ increases until $\lambda \approx D$, and then it decreases as $\lambda$ is further increased, for reasons discussed in the text.}\label{temp_dep}
\end{figure}

The phase-space argument~\cite{G1} for the bilayer fermionic systems indicates that there is a quadratic temperature
dependence ($\tau_D^{-1}\propto T^2$) at very low temperatures. We present the temperature dependence
of the drag rate scaled by $T^2$ for various coupling limits, as shown in Fig.~\ref{temp_dep}. In this figure, at weak
coupling limits (i.e., $\lambda=0.5,1,2,4$), the drag rate increases quadratically with the interaction strength
$\lambda$ as consistent with Eq.~\eqref{weak}. However, when the coupling strength is further increased, the
low-temperature drag rate decreases, as shown in Fig.~\ref{temp_dep} (see $\lambda=4,8$ and 16).

The upturn and peak in the scaled drag rates in Fig.~\ref{temp_dep} is due to the presence of acoustic plasmon modes,
which enhance the effective inter-layer interaction when they are thermally excited.  For weak coupling,
the plasmons have lower energy and therefore can be excited at lower temperatures.  As $\lambda$ increases,
the plasmon energies increase which results in a higher temperature in the upturn and peak of the drag rate.
After the peak, Landau damping reduces the effect of the plasmon enhancement, causing a decrease in the
scaled drag rate.

\subsection{Collective Modes}
In addition, we investigate the collective behavior of the system
at zero temperature to better understand their effects on mutual dipolar drag.  In our calculations, we use the
real part of the polarization function $\mbox{Re}\chi(Q,\Omega,T=0)$~\cite{Hu} at zero temperature because
at this limit the imaginary part vanishes, $\mbox{Im}\chi(Q,\Omega,T=0)=0$, outside the particle-hole continuum.

The dispersion for the collective modes is given by the zeros of $\varepsilon(q,\omega)$ which in the case of the equal bilayers is given by
\begin{equation}
[1 - V_{11}(q) \chi(q,\omega)]^2 - [V_{12}(q)\chi(q,\omega)]^2 = 0\, ,
\end{equation}
and it yields the solutions
\begin{equation}
\chi_{\pm}(Q,\Omega) = \frac{1}{V_{11} \pm V_{12}}.
\end{equation}

Solving the above equation for $\Omega$ at $T=0$ gives the collective mode dispersions (see Appendix B)
\begin{equation}
\Omega = \frac{Q(\tilde{V}_\pm + 1)}{\tilde{V}_{\pm}} \sqrt{Q^2
+ \frac{4 \tilde{V}_\pm^2}{2 \tilde{V}_\pm + 1}}\, ,\end{equation}
where $\tilde{V}_\pm=\lambda(\tilde{V}_{11}\pm\tilde{V}_{12})$.
This relation between $\Omega$ and $Q$ indicates that there are two separate collective modes above the
particle-hole continuum which is degenerate at small $Q$ and starts to be distinguished from each other
as $Q$ increases.

\begin{figure}[h]
\includegraphics[scale=0.6]{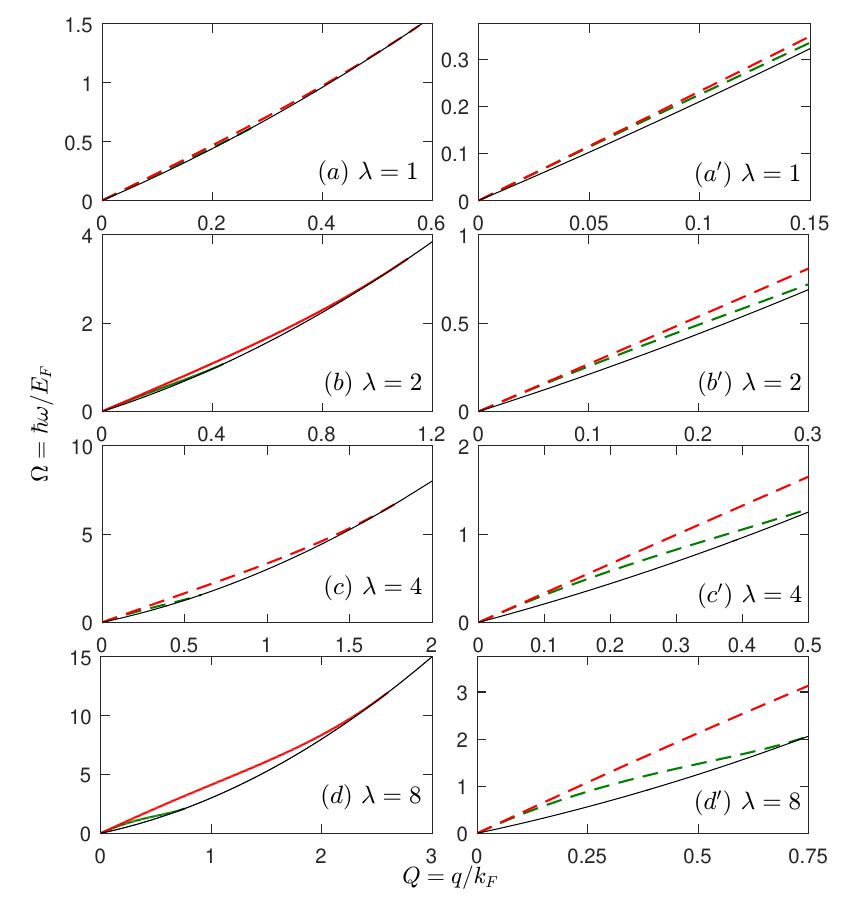}
\centering
\caption{(color online) The dimensionless frequency $\Omega$ at zero temperature is plotted as a function of $q/k_F$ when the layer separation distance is kept constant at $dk_F=2$. The plasmon dispersions are shown for four different values of the interaction strength $(a)\ \lambda=1$, $(b)\ \lambda=2$, $(c)\ \lambda=4$, and $(d)\ \lambda=8$. In each graph, the solid line represents the upper bound of the particle-hole continuum. In the area under this line, the collective modes enter the particle-hole continuum and Landau damping starts. The collective modes related to the charge-density oscillations in the layers of the system are demonstrated by the dashed lines in each figure. The zoomed-in view of the collective modes is demonstrated in the $(a')$, $(b')$, $(c')$, and $(d')$ graphs for the corresponding values of the interaction strength $\lambda$.}\label{collective}
\end{figure}

We also verify these results numerically and obtain the collective modes of the system at zero temperature as shown in Fig.~\ref{collective}. We show that for small values of $q$, the modes are well-defined and linear. As we increase the $q$ values, the collective modes split and then they disappear as a result of Landau damping when they merge with the particle-hole continuum.

\section{Discussion and Conclusion}
We investigate the transport properties of a bilayer 2D dipolar system of polarized fermions by calculating the drag rate
$\tau_D^{-1}$ as a function of temperature. To describe the correlation effects and screening within
a single layer, we use the Hubbard approximation for the effective intra-layer interaction.
The random-phase approximation (RPA) is adopted to obtain effective inter-layer interactions in the system.
We assume that there is no tunneling between the layers.

For sufficiently large
$\lambda$, there is a critical scaled separation distance $D_c(\lambda)$ below which
the system becomes unstable against the simultaneous formation of density
waves in both layers. This instability is caused by an enhancement in the
effective inter-layer interactions.  At a given $\lambda$, when the separation
between the layers $D \gg D_c(\lambda)$, the drag rate increases as $D^{-4}$ as the
separation is decreased.  However, as $D$ approaches $D_c$, the enhanced effective
inter-layer interaction causes the drag rate to increase faster than $D^{-4}$.  This
can be seen by comparing the panels with the same $\lambda$ in Figs. 2 and 3, where
$D = 1$ and $D=2$, respectively.  For $\lambda = 1$ and $2$ and $D_c \ll 1$ and
therefore the drag rate increases by approximately a factor of $2^4 = 16$ when $D$
is decreased from $2$ to $1$.  On the other hand for $\lambda = 4$ and $8$,  when
$D_c$ is close to $1$, the drag rate increases by more than a factor of $16$ when
$D$ decreases from $2$ to $1$.  This suggests that the distance dependence of drag
can be utilized experimentally to determine if a bilayer system is close to a
density wave instability.

In addition, the collective behavior of the system is studied at zero temperature where the imaginary part
of the polarization function is equal to zero, $\mbox{Im}\chi(Q,\Omega,T=0)=0$, outside the particle-hole continuum.
The modes are well-defined and linear for small $q$. Our analytic and numerical calculations show that the collective
modes split for increasing $q$ values up to a point and then they disappear as a result of Landau damping when they merge with the particle-hole continuum.

We believe that this model is applicable to studying the momentum transfer between the ultracold gases confined in
parallel layers.  For studies on ultracold gases, the crucial point of
this transport method will be the adjustment of the layer separation distance. Our calculations show that for weak
coupling limit $\lambda/D = a_0/d \ll1$, instability will not occur in the system. In addition, for an effective momentum
transfer, the layers must be positioned a few $a_0$'s apart
from each other because the drag rate decays rapidly with increasing layer separation distance.
For magnetic atomic species, $a_0$'s are of the order of tens of nanometers; for example, the calculated values of
$a_0$ for Cr, Er, and Dy are 2.4\,nm, 10.5\,nm, and 20.8\,nm, respectively. The
$a_0$'s for these species are much smaller than the typical trapping features in ultracold atom experiments.
However, the recent experiment by Lu {\it et al.}~\cite{Ketterle} demonstrated a 50\,nm interlayer separation by a
superresolution technique. 

While the experiment of Ref.\,\cite{Ketterle} provides an impetus for us to consider the bilayer geometry, direct comparison of the experiment with our results is impossible due to fundamental differences. The experiment features bosonic atoms in the BEC state, which has a completely different elementary excitation spectrum from the Fermi liquid considered here. Furthermore, the experiment has a prominent trap in the plane of the layers, thus most of the bilayer coupling is through the center of mass modes of the condensates. We hope that this work stimulates more interest in experimentally obtaining bilayer Fermi systems. We believe the superresolution trapping technique applied to a fermionic ultracold dipolar gas would be able to probe the physics described in this paper.
Another possibility would be the use of ultracold dipolar molecules for which $a_0$'s of the order of $10^{-6}$\,m, are easily obtained experimentally.

\begin{acknowledgments}
We would like to thank the Scientific and Technological Research Council
of Turkey (T\"{U}B\.{I}TAK Grant no:116F030) for its financial support. B.T. also thanks TUBA for their support.
B.Y.K.H. acknowledges the Professional Development Leave granted by the University of Akron and
partial support by T\"{U}B\.{I}TAK.
\end{acknowledgments}

\appendix
\section{Derivation of Drag Rate}
Following Rojo~\cite{Rojo} we calculate the rate of change in momentum of the dipoles in the second layer as a result of the scattering from dipoles in the first layer as
\begin{widetext}
\begin{equation}\label{momentum}
\begin{split}
\frac{d\textbf{P}_2}{dt} = \int \frac{d\textbf{q}}{(2\pi)^2}&\ \left(\hbar \textbf{q}\right)\ |U(q)|^2\int \frac{d\textbf{k}_1}{(2\pi)^2} \int \frac{d\textbf{k}_2}{(2\pi)^2}\, \ \delta(\epsilon_{\textbf{k}_1}+\epsilon_{\textbf{k}_2}-\epsilon_{\textbf{k}_1+\textbf{q}}-\epsilon_{\textbf{k}_2-\textbf{q}}) \\ \\
&\times\ [f_{\textbf{k}_1}(1-f_{\textbf{k}_1+\textbf{q}})f_{\textbf{k}_2}^0(1-f_{\textbf{k}_2-\textbf{q}}^0)
-f_{\textbf{k}_1+\textbf{q}}(1-f_{\textbf{k}_1})f_{\textbf{k}_2-\textbf{q}}^0(1-f_{\textbf{k}_2}^0)]\ \,.
\end{split}
\end{equation}
\end{widetext}
Eq.~\eqref{momentum} utilizes the Born approximation. The $\delta$-function enforces energy conservation during 
the scattering, the $(\hbar \textbf{q})\; |U(q)|^2$ gives the Born approximation momentum transfer rate, and the various
forms of the $f_{\textbf{k}_i}$-function arise from the probabilities of transitions from occupied states to empty ones.
Here, $f_{\textbf{k}_1}$ is the distribution function in layer 1 (the active layer), which is assumed to be drifted from the
equilibrium distribution $f^0$ by a small velocity 
$\textbf{v}_1$; {\it i.e.}, $f_{\textbf{k}_1} = f_{\textbf{k}_1-\frac{m\textbf v_1}{\hbar}}^0$.

To simplify the above expression, we make use of the following relations.

{(i)} Detailed balance condition for fermion systems:
\begin{equation}\label{detailed}\begin{split}
\Omega_{FF} & \doteq  f_{\textbf{k}_1}^0(1-f_{\textbf{k}_1+\textbf{q}}^0)f_{\textbf{k}_2}^0(1-f_{\textbf{k}_2-\textbf{q}}^0)  \\
            & = f_{\textbf{k}_1+\textbf{q}}^0(1-f_{\textbf{k}_1}^0)f_{\textbf{k}_2-\textbf{q}}^0(1-f_{\textbf{k}_2}^0)\,.
\end{split} \end{equation}

{(ii)} Linearization of $f_{\textbf{k}_1}$  and $f_{\textbf{k}_1+\textbf{q}}$ with respect to $\textbf{v}_1$:
\begin{equation}\label{f_one}\begin{split}
f_{\textbf{k}} &=f_{\textbf{k}-\frac{m\textbf v_1}{\hbar}}^0\approx f_{\textbf{k}}^0-\frac{\partial f_{\textbf{k}}^0}{\partial \epsilon_{\textbf{k}}}\hbar \textbf{k}\cdot \textbf v_1 \\
          &  \equiv f_{\textbf{k}}^0 -\frac{1}{k_BT}f_{\textbf{k}}^0(1-f_{\textbf{k}}^0)\hbar \textbf{k}\cdot \textbf v_1\,
\end{split}\end{equation}

We can rewrite the last term in Eq.~\eqref{momentum} by substituting in the linearized expressions for $f_{\textbf{k}}$, ignoring the nonlinear $\textbf{v}_1$ terms and using the detailed balance condition, Eq.~\eqref{detailed}, to give
\begin{equation}\label{curly4}
\begin{split}
f_{\textbf{k}_1}&(1-f_{\textbf{k}_1+\textbf{q}})f_{\textbf{k}_2}^0(1-f_{\textbf{k}_2-\textbf{q}}^0)\\
                & -f_{\textbf{k}_1+\textbf{q}}(1-f_{\textbf{k}_1})f_{\textbf{k}_2-\textbf{q}}^0(1-f_{\textbf{k}_2}^0))\\
 =&\frac{1}{k_BT}\ \hbar\ (\textbf{q}\cdot \textbf{v}_1)\ \Omega_{FF} \\
           =&\frac{1}{k_BT}\ \hbar\ (\textbf{q}\cdot\textbf{v}_1) \left\{f_{\textbf{k}_1}^0(1-f_{\textbf{k}_1+\textbf{q}}^0)f_{\textbf{k}_2}^0(1-f_{\textbf{k}_2-\textbf{q}}^0)\right\}\,,
\end{split}
\end{equation}

The rate of the momentum change becomes
\begin{widetext}
\begin{equation}\label{P}
\frac{d\textbf{P}_2}{dt} = \frac{\textbf{v}_1}{2k_BT} \int \frac{d\textbf{q}}{(2\pi)^2}(\hbar q)^2\ |U(q)|^2\int \frac{d\textbf{k}_1}{(2\pi)^2} \int \frac{d\textbf{k}_2}{(2\pi)^2} \left\{f_{\textbf{k}_1}^0(1-f_{\textbf{k}_1+\textbf{q}}^0)f_{\textbf{k}_2}^0(1-f_{\textbf{k}_2-\textbf{q}}^0)\right\} \delta(\epsilon_{\textbf{k}_1}+\epsilon_{\textbf{k}_2}-\epsilon_{\textbf{k}_1+\textbf{q}}-\epsilon_{\textbf{k}_2-\textbf{q}}) \,.
\end{equation}
\end{widetext}
At this point, we use the following relations,
\begin{equation}\label{delta}\begin{split}
       \delta(\epsilon_{\textbf{k}_1}&+\epsilon_{\textbf{k}_2}-\epsilon_{\textbf{k}_1+\textbf{q}}-\epsilon_{\textbf{k}_2-\textbf{q}})=  \\
       & \hbar \int_{-\infty}^{\infty}  d\omega \ \delta(\hbar\omega-\epsilon_{k_1}+\epsilon_{k_1+q}) \delta(\hbar\omega+\epsilon_{k_2}-\epsilon_{k_2-q}),
\end{split} \end{equation}
\begin{equation}\label{equiv}
        f^0(\epsilon_\textbf{k})\left[1-f^0(\epsilon_\textbf{k}+\hbar\omega)\right]=
        \frac{\left[f^0(\epsilon_\textbf{k})-f^0(\epsilon_\textbf{k}+\hbar\omega)\right]}{\left[1-\exp(-\hbar\omega/k_BT)\right]}\,
\end{equation}
\\
\begin{equation}\label{rel}\begin{split}
n_B(\hbar \omega)n_B(-\hbar \omega)=-\frac{1}{4\sinh^2(\hbar\omega/2k_BT)}.
\end{split} \end{equation}
\\
We also introduce the polarization function $\chi^0(q,\omega)$
\begin{widetext}
\begin{equation}\label{chi}
\int \frac{d\textbf{k}_1}{(2\pi)^2}\ (f_{\textbf{k}_1}^0-f_{\textbf{k}_1+\textbf{q}}^0)\ \delta(\hbar\omega-\epsilon_{\textbf{k}_1}+\epsilon_{\textbf{k}_1+\textbf{q}})=
\frac{1}{\pi}\mbox{Im}\int\frac{d\textbf{k}_1}{(2\pi)^2}\ \frac{f_{\textbf{k}_1}^0-f_{\textbf{k}_1+\textbf{q}}^0}{(\hbar\omega-\epsilon_{\textbf{k}_1}+\epsilon_{\textbf{k}_1+\textbf{q}}-\mbox{\textbf{i}}\eta)}\equiv\mbox{Im}\chi_1^0(q,\omega)\,. \end{equation}
\end{widetext}

Using equations~\eqref{delta}-\eqref{chi} into Eq.~\eqref{P} and performing some variable changes yields
\begin{equation}\label{result1}\begin{split}
\frac{dP_2}{dt} = \frac{v_1\hbar^2}{8\pi^2k_BT}\ \int & dq \ q^3\ |U(q)|^2\ \\
 &\int d\omega \frac{\mbox{Im}\chi_1^0(q,\omega)\mbox{Im}\chi_2^0(q,\omega)}{\sinh^2(\hbar\omega/2k_BT)}
\end{split} \end{equation}
Finally, introducing the momentum per particle in the second layer, $p_2=P_2/n_2$ and writing the momentum per particle in the first layer as $p_1=m_1v_1$, we obtain the rate of momentum transfer between the layers as
\begin{widetext}
\begin{equation}\label{result2}
\tau_D^{-1} =\frac{\hbar^2 }{8m_1n_2k_BT\pi^2} \int  dq \ q^3\ |U(q)|^2\ \int d\omega \frac{\mbox{Im}\chi_1^0(q,\omega)\mbox{Im}\chi_2^0(q,\omega)}{\sinh^2(\hbar\omega/2k_BT)}\, .
\end{equation}
\end{widetext}
\section{Collective mode dispersions}
To find the collective mode dispersions at $T=0$, we solve
\begin{equation}
\chi_{\pm}(Q,\Omega) = \frac{1}{V_{11} \pm V_{12}},
\end{equation}
using the dimensionless quantities $\tilde \chi$, $\tilde{V}_{11}$ and $\tilde{V}_{12}$.
We look for collective modes above the particle hole continuum, so that
\begin{subequations}
\begin{align}
2\tilde\chi(Q,\Omega) &= -1 +\frac1Q\left[ \sqrt{a_+^2 - 1} - \sqrt{a_-^2 - 1}\right]\\
a_\pm &= \frac12\left(\frac{\Omega}Q \pm Q\right).
\end{align}
\end{subequations}
Thus, the collective modes are given by
\begin{equation}
\sqrt{a_+^2 - 1} - \sqrt{a_-^2 - 1} = Q\left[1 + \frac{1}{\lambda(\tilde{V}_{11} \pm \tilde{V}_{12})}\right].  \label{eq:A5}
\end{equation}
Let us define the right hand side of Eq.~(\ref{eq:A5}) as
\begin{equation}
P_\pm \stackrel{\mathrm{def.}}{=}\  Q\left(1 + \frac1{\tilde{V}_{\pm}}\right) = Q\left(\frac{\tilde{V}_\pm+1}{\tilde{V}_\pm}\right).\label{eq:A6a}
\end{equation}
where $\tilde{V}_{\pm} = \lambda(\tilde{V}_{11} \pm \tilde{V}_{12})$.
Squaring both sides of Eq.~(\ref{eq:A5}) and defining
\begin{equation}
\eta = \frac14\left(\frac{\Omega^2}{Q^2} + Q^2\right) - 1 \label{eq:def_eta}
\end{equation}
gives
\begin{equation}
- 2 \sqrt{\eta^2 - \frac{\Omega^2}4} = P_\pm^2 - 2\eta.
\end{equation}
After some manipulations, we obtain
\begin{equation}\label{omeg}
\Omega = P_\pm Q \sqrt{1 + \frac{4}{P_\pm^2 - Q^2}}
\end{equation}
Substituting the expression for $P_\pm$ in Eq.~(\ref{eq:A6a}) into this gives the collective mode dispersions
\begin{equation}
\Omega = \frac{Q(\tilde{V}_\pm + 1)}{\tilde{V}_{\pm}} \sqrt{Q^2 + \frac{4 \tilde{V}_\pm^2}{2 \tilde{V}_\pm + 1}}\,.\end{equation}


\begin{thebibliography}{99}
\bibitem{attractint} In the case of attractive inter-layer interactions ({\em e.g.}, parallel dipoles), 
when the inter-layer pairing is weak and the Born approximation for the inter-layer scattering is valid, 
the results are the same as the ones presented here for the repulsive case.
\end{thebibliography}
\end{document}